\documentstyle[twocolumn,aps,psfig]{revtex}

\begin{document}
\draft

\twocolumn[\hsize\textwidth\columnwidth\hsize\csname @twocolumnfalse\endcsname

\title{
Stripes due to the next-nearest neighbor exchange in high-$T_{\rm c}$ 
cuprates
}

\author{T\^oru Sakai}
\address{
Faculty of Science, Himeji Institute of Technology, 3-2-1 Kouto, Kamigori,
Ako-gun, Hyogo 678-1297, Japan
}

\date{February 2001}
\maketitle 

\begin{abstract}
We propose  
a possible mechanism of the charge stripe order due to the
next-nearest neighbor exchange interaction $J'$ in the
two-dimensional $t$-$J$
model, based on the concept of the phase separation.
We also calculate some hole correlation functions of the finite
cluster of the model using the numerical diagonalization,
to examine the realization of the mechanism.
It is also found that the next-nearest neighbor hopping $t'$ 
suppresses the stripe order induced by the present mechanism 
for $t'<0$, while it enhances for $t'>0$.  
\end{abstract}

\vskip2pc]
\narrowtext

%
%

The charge stripe order\cite{tranquada1,tranquada2} observed
in the high-temperature cuprates
superconductors is one of the most interesting current topics
on the strongly correlated electron systems.
In particular since the discovery of the coexistence
with the superconductivity in La$_{1.6-x}$Nd$_{0.4}$Sr$_x$CuO$_4$
\cite{tranquada3},
the mechanism of the stripe formation has been studied in many works.
The numerical study\cite{white} based on
the density matrix renormalization group
suggested that such a stripe phase can appear in the two-dimensional
$t$-$J$ model. 
On the other hand, the numerical diagonalization of the 
$4\times 4$ $t$-$J$ cluster with two holes\cite{manousakis} 
indicated that 
the stripe order occurs only in some low-lying excited states, 
rather than the ground state. 
The realization of the stripe order in the simple $t$-$J$ model 
is still an open problem. 

It is well known that the $t$-$J$ model should exhibit the phase
separation for sufficiently large $J/t$.\cite{PS}
The high temperature expansion suggested such a state is realized
for $J/t \geq 1$.\cite{high}
Some small cluster calculations have shown
that a larger cluster of the holes is stable rather than a pair
even in more realistic parameter region ($J/t\geq 0.5$).\cite{didier}
In the present paper,
we propose a possible mechanism of the stripe order formation
due to the additional next-nearest-neighbor exchange interaction $J'$
based on a naive argument valid in the phase separation region
of the $t$-$J$ model.
Since the next-nearest-neighbor hopping $t'$ has been
revealed to be quite large for Sr$_2$CuO$_2$Cl$_2$ ($t'\sim 0.3t$)
\cite{tohyama},
$J'$ is also expected to be finite in some real cuprates.
Thus we consider the square-lattice $t$-$t'$-$J$-$J'$ model 
to discuss on the mechanism of the stripe. 
We also calculate the three- and four-hole correlation functions
of the $4\times 4$ cluster with four holes, to examine the
realization of the mechanism.

%
%

We consider the two-dimensional $t$-$J$ model in the presence of the
next-nearest-neighbor hopping $t'$ and the exchange interaction $J'$.
The Hamiltonian is given by the form 
\begin{eqnarray}
\label{ham}
H = &- t \sum_{<\bf i,\bf j>, \sigma}
            ({c}_{\bf j,\sigma}^\dagger {c}_{\bf i,\sigma}
           + {c}_{\bf i,\sigma}^\dagger {c}_{\bf j,\sigma} )
\nonumber
\\
 &- t' \sum_{<\bf i,\bf j>', \sigma}
            ({c}_{\bf j,\sigma}^\dagger {c}_{\bf i,\sigma}
           + {c}_{\bf i,\sigma}^\dagger {c}_{\bf j,\sigma} )
\nonumber
\\
  &+ J \sum_{<\bf i,\bf j>}
           ( {\bf S}_{\bf i} \cdot {\bf S}_{\bf j}
           - \textstyle{1 \over 4} n_{\bf i} n_{\bf j} ) \nonumber
\nonumber
\\
  &+ J' \sum_{<\bf i,\bf j>'}
           ( {\bf S}_{\bf i} \cdot {\bf S}_{\bf j}
           - \textstyle{1 \over 4} n_{\bf i} n_{\bf j} )
\end{eqnarray}
where $\sum_{<\bf i,\bf j>}$ and 
$\sum_{<\bf i,\bf j>'}$
mean the summation over all the nearest-neighbor and 
the next-nearest-neighbor sites, respectively.
Throughout the paper, all the energies are
measured in units of $t$.
%
%
We assume the next-nearest-neighbor exchange interaction is 
antiferromagnetic ($J'>0$), as was revealed for La$_2$CuO$_4$ 
by the theoretical study based on the {\it ab initio} 
calculation. \cite{annett}
The antiferromagnetic $J'$ term can also be derived from 
the strong correlation expansion of the Hubbard Hamiltonian 
up to the order of $t^4/U^3$. \cite{coldea} 
Since $t'$ plays no essential roles in the following argument, 
we set $t'=0$ at first. 

Consider the naive argument to explain the hole pairing due to
the antiferromagnetic short range order: a pair of holes sitting
on the adjacent cites is more stable than two separated holes,
because the former breaks 7 $J$ bonds, while the latter 8 $J$ bonds.
Following the argument,
larger hole clusters are expected to be formed for sufficiently
large $J$.
In such a situation we consider the effect of $J'$.
(We assume $J'$ is not so large that the antiferromagnetic short
range order is completely broken.)
At first we compare the stability of three-hole cluster in two
different shapes, shown in Figs. 1(a) and (b), respectively.
The number of $J$ bonds are the same between them, but
(a) has one more broken $J'$ bond than (b).
When the antiferromagnetic short range correlation is developed,
the $J$ bond should lead to the advantage of the energy, while the $J'$
to the disadvantage, as far as $J$ and $J'$ are antiferromagnetic.
Then (a) is expected to be more stable than (b).
Thus the three hole cluster should prefer the line shape like (a)
to the corner shape like (b).
Next we consider the four-hole cluster with the two shapes, shown in
Figs.1 (c) and (d), respectively.
In this case the number of $J$ bonds is also different.
One more $J$ bond and two more $J'$ bonds are broken in the shape (c)
than (d).
Assuming that the antiferromagnetic short range order is so large
that the next-nearest-neighbor spin correlation is almost the same
as the next one in amplitude,
the line shape (c) is more preferable than (d) under the
condition $J'\geq J/2$.
This condition is easily revealed to be approximately valid
in comparison between the line-shaped and the square-shaped
larger clusters with the same number of holes.
Thus large line-shaped clusters of holes should be
formed for sufficiently large $J'$.
This naive argument is expected to give a possible mechanism of
the charge stripe order.

%
%
\begin{figure}[htb]
\begin{center}
\mbox{\psfig{figure=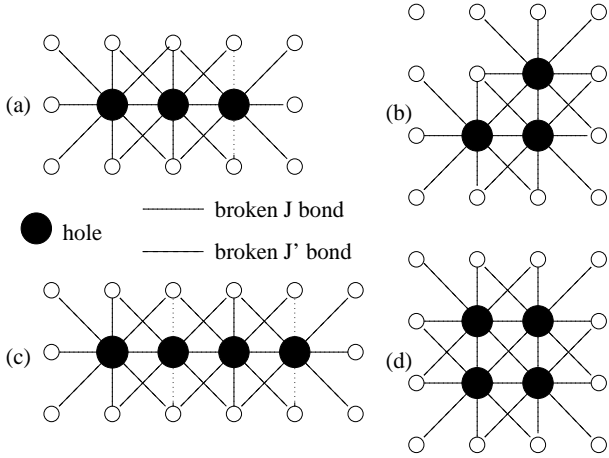,width=8cm,height=6cm,angle=0}}
\end{center}
\caption{
Schematic figures to discuss on the stability of
the three-hole and four-hole clusters.
\label{fig1}
}
\end{figure}
%
%
\begin{figure}[htb]
\begin{center}
\mbox{\psfig{figure=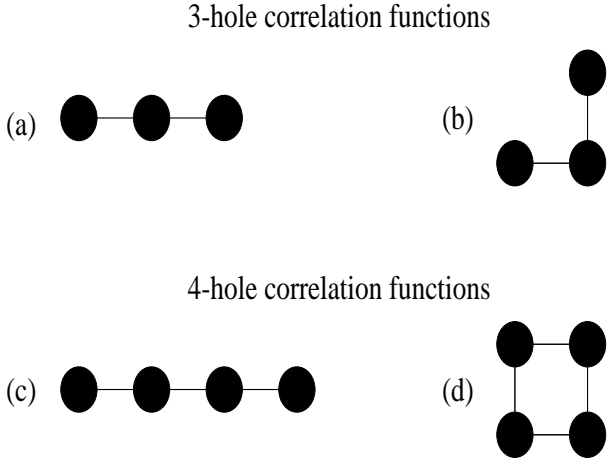,width=8cm,height=6cm,angle=0}}
\end{center}
\caption{
Configurations of the many-hole correlation functions;
(a)$C_{\rm St}^{(3)}$, (b)$C_{\rm PS}^{(3)}$,
(c)$C_{\rm St}^{(4)}$ and (d)$C_{\rm PS}^{(4)}$.
\label{fig2}
}
\end{figure}

%
%

In order to examine the realization of the mechanism of
the charge stripe order discussed in the previous section,
we calculate the three- and four-hole correlation functions
defined as
\begin{eqnarray}
\label{hhc}
C_{\rm St}^{(3)}=\big< \sum_{\bf i}n^h_{\bf i}n^h_{{\bf i}+{\hat x}}
n^h_{{\bf i}+2{\hat x}}\big>
\\
C_{\rm PS}^{(3)}=\big< \sum_{\bf i}n^h_{\bf i}n^h_{{\bf i}+{\hat x}}
n^h_{{\bf i}+{\hat x}+{\hat y}}\big>
\\
C_{\rm St}^{(4)}=\big< \sum_{\bf i}n^h_{\bf i}n^h_{{\bf i}+{\hat x}}
n^h_{{\bf i}+2{\hat x}}n^h_{{\bf i}+3{\hat x}}\big>
\\
C_{\rm PS}^{(4)}=\big< \sum_{\bf i}n^h_{\bf i}n^h_{{\bf i}+{\hat x}}
n^h_{{\bf i}+{\hat y}}n^h_{{\bf i}+{\hat x}+{\hat y}}\big>,
\end{eqnarray}
in the ground state of the finite cluster $t$-$t'$-$J$-$J'$ model 
($t'=0$).
$C_{\rm St}^{(3)}$ and $C_{\rm St}^{(4)}$ are supposed to
represent a relative strength of the stripe order,
while $C_{\rm PS}^{(3)}$ and $C_{\rm PS}^{(4)}$ measure a
tendency towards the ordinary phase separation.
They are calculated for the $4\times 4$ cluster with four holes,
for which the ground state has the $d$-wave like rotational symmetry
for $J \geq 0.3$.\cite{didier}
(We neglect the other ground states which appear in smaller $J$ 
regions for simplicity.) 
The calculated three- and four-hole correlation functions are plotted
versus $J'$ with fixed $J$ (=0.6 and 0.8), in Figs. 3 and 4,
respectively.
We detected a first-order transition (a level cross) at some
critical value $J'_c$ ($J'_c$ depends on $J$.)
and found that the line-shaped correlation
is larger than the square-shaped one for $J'\geq J'_c$,
while it is reversed for $J'\leq J'_c$ in both
Figs. 3 and 4.
It implies that
the charge stripe order is possibly realized in the bulk system
for sufficiently large $J'$, in agreement with the mechanism
proposed in the previous section.
Then $J'_c$ is expected to be the boundary between
the phase separation and the stripe ordered phases
in the thermodynamic limit.
Plotting the calculated $J'_c$ for various values of $J$,  
we give a phase diagram in the $J'$-$J$ plane for $t'=0$ 
(solid circles) in Fig. 5.
We can also understand that the excited state with the stripe 
order, which was found in the previous numerical 
study\cite{manousakis}, is stabilized by the next-nearest-neighbor 
exchange interaction in the upper phase in Fig. 5. 
%
%
\begin{figure}[htb]
\begin{center}
\mbox{\psfig{figure=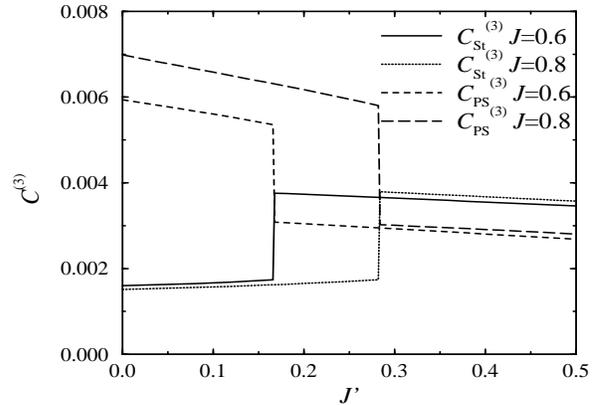,width=8cm,height=6cm,angle=0}}
\end{center}
\caption{
Three-hole correlation functions versus $J'$ with fixed $J$(=0.6 and
0.8).
\label{fig3}
}
\end{figure}

%
%
\begin{figure}[htb]
\begin{center}
\mbox{\psfig{figure=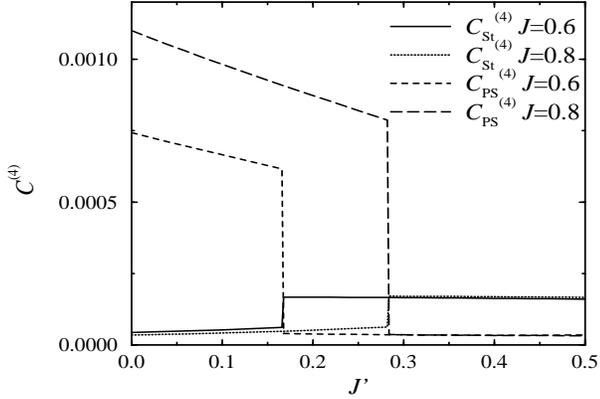,width=8cm,height=6cm,angle=0}}
\end{center}
\caption{
Four-hole correlation functions versus $J'$ with fixed $J$(=0.6 and
0.8).
\label{fig4}
}
\end{figure}

%
%
\begin{figure}[htb]
\begin{center}
\mbox{\psfig{figure=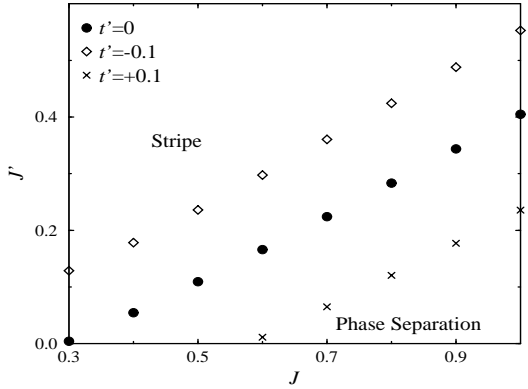,width=8cm,height=6cm,angle=-90}}
\end{center}
\caption{
Phase diagrams in $J'$-$J$ plane for $t'$=0, -0.1 and 0.1.
\label{fig5}
}
\end{figure}

%
%

The phase diagram for $t'=0$ in Fig. 5 indicates an interesting point
that the stripe order is possibly realized even if $J'$ is much
smaller than $J/2$ in small-$J$ region around $J\sim 0.4$,
which is realistic for the high-$T_c$ cuprates.
%
%
Some recent theoretical analyses\cite{gimm,hellberg,pryadko}
 on the simple $t$-$J$ model 
actually revealed that 
the phase separation occurs even in such a realistic parameter 
region. 
The present result of the phase separation-stripe boundary 
$J_{\rm c}\sim 0.3$ for $J'=t'=0$ 
in Fig. 5 is consisent with these results.  
It implies that the scenario of the stripe formation based on 
the next-nearest neighbor exchange interaction is possibly 
valid for real cuprates, 
although the presice phase boundary is still controversial. 
Note that the present anarysis does not distinguish the static 
stripe order and the dynamical one like the 
charge strings, which was predicted by the phonon-induced polaron 
mechanism.\cite{kusmartsev}  
It would be an interesting futre work to study on 
such a dynamical stripe, which may give some hints to explain the
coexistence of the stripe order and the superconductivity
observed in La$_{1.6-x}$Nd$_{0.4}$Sr$_x$CuO$_4$. 

Finally, 
we consider the effect of the next-nearest-neighbor hopping 
$t'$ in the present mechanism of the stripe formation. 
For this purpose, 
the phase boundaries between the stripe and the pairing (or 
phase separation) phases for $t'$=-0.1 (diamonds) and $t'$=0.1 
(crosses) are shown in Fig. 5. 
The negative and positive $t'$ are corresponding to hole and 
electron doping cases, respectively. 
The phase diagram suggests that 
the negative $t'$ suppresses the stripes, while the positive 
$t'$ enhances it. 
The result agrees with the numerical studies\cite{white2,tohyama2}  
at least for small $t'$, although they didn't consider $J'$. 
It implies that the stripe due to $J'$ in the present mechanism 
has the same feature as the one which was investigated in those 
previous works. 
Actually Fig. 5 indicates that the stripe can occur even for 
$J'=0$ at least in the case of the positive $t'$. 
It would be more interesting to perform the same calculation for 
more realistic hole density, close to 1/8, if possible. 
(For example, the 32-site cluster with 4holes is desirable, 
but it is difficult for the present computer systems.) 

%
%
The recent high-resolution inelastic neutron scattering experiment 
\cite{coldea} indicated that the ring (four-spin) exchange interaction is 
more important to explain the observed spin-wave 
dispersion of La$_2$CuO$_4$, rather than the next-nearest-neighbor 
exchange interation. 
Thus we should also take the ring exchange interaction into 
account for more quantitative study. 

In summary, 
we proposed a possible mechanism of the charge stripe formation 
based on the next-nearest-neighbor exchange interaction $J'$ 
in the high $T_{\rm c}$ cuprates. 
The many-hole correlation functions of the $4\times 4$ lattice 
$t$-$t'$-$J$-$J'$ model indicated that even small $J'$ 
possibly induces the stripe order for realistic values of $J$. 
In addition the next-nearest-neighbor hopping $t'$ was 
revealed to suppress the stripe for $t'<0$, but enhance it 
for $t'>0$.

%
%
We thank D. Poilblanc and T. M. Rice for 
fruitful discussions. 
The computation in this work has been done using the
facilities of the Supercomputer Center, Institute for Solid
State Physics, University of Tokyo.
This research was supported in part by Grant-in-Aid 
for the Scientific Research Fund from the Ministry 
of Education, Science, Sports and Culture (11440103).

\end{document}